\documentclass[reprint,a4paper,amsmath,aps,prd,twocolumn,superscriptaddress,preprintnumbers,noshowpacs]{revtex4}
\usepackage{CJK}
\usepackage{amsmath}
\usepackage{enumerate}
\usepackage{amssymb}
\usepackage{amsfonts}
\usepackage{mathrsfs}
\usepackage{latexsym}
\usepackage{bm}
\usepackage{amscd}
\usepackage{graphicx}
\usepackage{epsfig}
\usepackage{hhline,multirow}
\usepackage{dcolumn}
\usepackage{url}
\usepackage[dvipsnames]{xcolor}
\definecolor{DarkBlue}{rgb}{0.0, 0.0, 0.4}
\usepackage[colorlinks=true,linkcolor=blue,urlcolor=DarkBlue]{hyperref}
\usepackage{indentfirst}
\usepackage{subfigure}
\usepackage{psfrag}
\usepackage{slashed}
\usepackage{float}
\usepackage{setspace}

\newcommand{\mpi}{M_\pi}

\begin{document}
\preprint{ADP-21-6/T1153}

\title{Chiral extrapolation of the charged-pion magnetic polarizability with Pad\'e approximant}
\author{Fangcheng He}
\affiliation{Institute of High Energy Physics, CAS,
	Beijing 100049, China}
\affiliation{CAS Key Laboratory of Theoretical Physics, Institute of Theoretical Physics, CAS, Beijing 100190, China}
\author{Derek B. Leinweber}
\affiliation{Centre for the Subatomic Structure of Matter (CSSM), Department of Physics, University of Adelaide,
	Adelaide, South Australia 5005, Australia}
\author{Anthony W. Thomas}
\affiliation{Centre for the Subatomic Structure of Matter (CSSM), Department of Physics, University of Adelaide,
	Adelaide, South Australia 5005, Australia}
\author{Ping Wang}
\affiliation{Institute of High Energy Physics, CAS,
	Beijing 100049, China}

\begin{abstract}
The background magnetic-field formalism of lattice QCD has been used recently to calculate the
magnetic polarizability of the charged pion.  These $n_f = 2 + 1$ numerical simulations are
electroquenched, such that the virtual sea-quarks of the QCD vacuum do not interact with the
background field.  To understand the impact of this, we draw on partially quenched chiral
perturbation theory.  In this case, the leading term proportional to $1/M_\pi$ arises at tree level
from $\mathcal{L}_4$.  To describe the results from lattice QCD, while maintaining the exact
leading terms of chiral perturbation theory, we introduce a Pad\'e approximant designed to
reproduce the slow variation observed in the lattice QCD results.  Two-loop contributions are
introduced to assess the systematic uncertainty associated with higher-order terms of the
expansion.  Upon extrapolation, the magnetic polarizability of the charged pion at the physical
pion mass is found to be $\beta_{\pi^\pm}=-1.70\,(14)_{\rm stat}(25)_{\rm syst}\times 10^{-4}$
fm$^3$, in good agreement with the recent experimental measurement.
\end{abstract}


\maketitle

\section{Introduction}

Determining the electromagnetic structure of baryons and mesons presents a contemporary challenge
of broad interest in hadron physics.  The internal structure of hadrons is governed by the
interactions between quarks and gluons described by quantum chromodynamics (QCD). Although QCD is
well established to describe the strong interactions, it is very difficult to study hadronic
physics using QCD directly, due to its nonperturbative behavior.  

Many phenomenological models as well as effective field theory have been utilized to learn about
the mechanisms at play in determining the hadron spectrum, hadron structure and hadronic
interactions.
The most rigorous way to study hadron physics is through lattice QCD. It is based on the first
principles of the quantum field theory and provides an avenue for the {\it ab initio} calculation
of Green functions via simulations of the path integral in a discrete space-time lattice. 

The electromagnetic polarizability is a fundamental property characterizing the structure of a
hadron. The observable reflects a dynamical response of a hadron to an external electromagnetic
probe.  As the lightest meson, the pion polarizability is of special interest. It is very difficult
to measure accurately, due to the short lifetime of the pion \cite{Ahrens:2004mg,Adolph:2014kgj}.  A
relatively recent measurement by the COMPASS collaboration \cite{Adolph:2014kgj} provides
$\beta_{\pi^\pm}=(-2.0\pm 0.6 \pm 0.7)\times 10^{-4}$ fm$^3$ for the charged pion.  
Here the uncertainties are statistical and systematic respectively.

This experimental measurement is complemented by theoretical calculations based on models, such as
the quark confinement model \cite{Ivanov:1991kw}, Nambu-Jona-Lasinio model 
\cite{Bernard:1988wi,Dorokhov:1997rv}, linear sigma model \cite{Bernard:1988gp}, dispersion sum
rules \cite{Filkov:1982cx,Donoghue:1993kw}, as well as chiral perturbation theory
\cite{Burgi:1996qi,Gasser:2006qa}.

In this investigation we analyze lattice QCD results for the magnetic polarizability of the charged
pion determined via the uniform background magnetic field formalism.  Early work with the
background-field approach calculated baryon magnetic moments
\cite{Martinelli:1982cb,Bernard:1982yu}.  The first attempt to calculate a polarizability with the
background field method was made by Fiebig {\it et al.} \cite{Fiebig:1988en}.  The formalism for
calculating the magnetic polarizability of a baryon within the background field method was outlined
in Ref.~\cite{Burkardt:1996vb}.

Today, there are several calculations of light-hadron magnetic polarizabilities using the
background-field formalism
\cite{%
Lee:2005dq,%
Primer:2013pva,
Luschevskaya:2014lga,
Luschevskaya:2015cko,
Bali:2017ian,%
Bignell:2019vpy,
Ding:2020hxw%
},
with advances in algorithms complementing increased supercomputing resources to
significantly reduce both systematic and statistical errors over time.
In addition, first attempts to calculate the polarizability of light nuclei \cite{Chang:2015qxa}
have been made.
More recently, Landau and Laplacian $SU(3)\times U(1)$ projection methods have been created to
isolate the ground states of hadrons in an external magnetic field
\cite{Bignell:2018acn,Bignell:2020xkf,Bignell:2020dze}.

While the chiral extrapolation of the magnetic polarizability of the nucleon and neutral pion has been
considered \cite{Hall:2013dva, Bignell:2018acn,Bignell:2020xkf,He:2020ysm}, a chiral extrapolation
of lattice QCD results for the charged-pion magnetic polarizability $\beta_{\pi^\pm}$ remains.  

In this paper, we will extrapolate the lattice results of Ref.~\cite{Bignell:2020dze} for
$\beta_{\pi^\pm}$ to the physical pion mass. The results of Ref.~\cite{Bignell:2020dze} employ a
new Laplacian-mode projection technique that isolates the state of interest and enables accurate
determinations of the small energy shifts induced by the background magnetic field. We will analyze
the one loop diagrams from partially quenched chiral perturbation theory to identify the leading
contributions of quark-flow connected and disconnected diagrams separately.

\section{Partially Quenched $\mathbf\chi$PT}

A naive approach to chiral extrapolation is to simply use low-order polynomial fit functions of the
quark mass to fit the lattice results. However, such a procedure is not correct as it neglects the
effects of the meson cloud, which can produce terms nonanalytic in the quark mass.  These terms can
generate rapid variation in observables for pion masses below 400 to 500 MeV
\cite{Thomas:2002sj}. The nonanalytic terms are crucial in obtaining the correct extrapolated
results at the physical pion mass.  

Chiral perturbation theory ($\chi$PT) provides a robust framework for determining the nonanalytic
terms and their coefficients.  The coefficients are model-independent and should not be taken as
fit parameters.
This approach has been used to extrapolate many hadronic observables
\cite{Leinweber:2003dg,Leinweber:2005xz,%
Allton:2005fb,Armour:2005mk,%
Young:2004tb,Leinweber:2004tc,Wang:1900ta,%
Leinweber:2006ug,Hall:2013oga}.

For the magnetic polarizability of the charged pion, the tree level contribution starting from the
next-to-leading order Lagrangian $\mathcal{L}_4$ provides a leading term to $\beta_{\pi^\pm}$ of
order $1/\mpi$ with a well-determined coefficient \cite{Burgi:1996qi}.  The leading nonanalytic
contribution to the Compton amplitude proportional to ${\rm log}(m_\pi)$ has its origin in the
two-loop diagrams of $\chi$PT \cite{Gasser:2006qa}.

We find that the lattice QCD results for $\beta_{\pi^\pm}$ are described very well over the
available pion-mass range by a Pad\'e approximant involving three terms. This approximation
provides an interpolation between the light quark-mass regime where $\chi$PT is robust to the
larger quark-mass regime where the lattice-QCD results are smooth and slowly varying as a function
of the quark mass.  A similar approach was explored in Ref.~\cite{HackettJones:2000qk} where baryon
magnetic moments were extrapolated to the physical point.

To explore the systematic errors of the approach and the importance of higher-order terms in the
chiral expansion, two-loop contributions are also considered \cite{Gasser:2006qa}.  These
contributions are found to be small relative to the leading contributions.  With this
consideration, the magnetic polarizability of the charged pion at the physical point is
$\beta_{\pi^\pm}=-1.70\,(14)_{\rm stat}(25)_{\rm syst}\times 10^{-4}$ fm$^3$, in good agreement
with the recent experimental measurement of Ref.~\cite{Adolph:2014kgj}.

\begin{figure}[t]
\begin{center}
\includegraphics[scale=0.85]{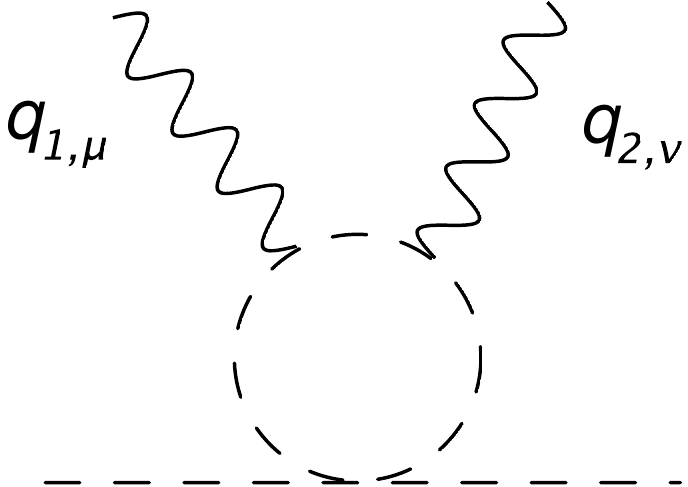}
\caption{The leading one-loop diagram for the pion magnetic polarizability}.
\label{fig:Nfig1}
\end{center}
\end{figure}

For pion-photon scattering, the Taylor expansion of the Compton amplitude in photon energies at
threshold can be expressed as
\begin{eqnarray}
\label{mag}
T&=&-2\, \left [ \vec{\epsilon}_1\cdot\vec{\epsilon}_2^{\,*}\, (e^2 - 4\pi\,
  M_\pi\alpha_\pi \, \omega_1 \, \omega_2)- \right . \nonumber \\
&&\qquad \left . 4\pi \,
  M_\pi \, \beta_\pi\, ( \vec{q}_1\times\vec{\epsilon}_1) \cdot
  (\vec{q}_2\times\vec{\epsilon}_2^{\,*})+ \cdots \right ] \, ,
\end{eqnarray}
where $\alpha_\pi$ and $\beta_\pi$ are called the electromagnetic polarizabilities.  There have been
several calculations addressing the pion electromagnetic polarizability within chiral effective
field theory \cite{Bijnens:1987dc,Burgi:1996qi,Gasser:2006qa}.  The chiral Lagrangian is organized
in the following terms
\begin{eqnarray}
\mathcal{L}=\mathcal{L}_2+\mathcal{L}_4+\mathcal{L}_6+\cdots \, ,
\end{eqnarray}
where the subscripts refer to the chiral order.  For the one-loop diagrams, only the Lagrangian at
leading order $\mathcal{L}_2$ is used.  The expression for $\mathcal{L}_2$ is
\begin{eqnarray}
\mathcal{L}_2=\frac{f_\pi^2}{4}\, Tr[D_\mu UD^\mu U^\dag] + \frac{f_\pi^2}{4}\, Tr[m(U+U^\dag)] \, ,
\end{eqnarray}
where $U=e^{2i\phi/f_\pi}$ and $f_\pi=92.4(3)$\,MeV is the pion decay constant \cite{Groom:2000in}.
$\phi$ is the matrix of pseudoscalar fields
\begin{eqnarray}
\phi=\frac1{\sqrt{2}}\left(                
\begin{array}{lcr}                       
\frac1{\sqrt{2}}\pi^0+\frac1{\sqrt{6}}\eta & \pi^+ & K^+ \\    
~~\pi^- & -\frac1{\sqrt{2}}\pi^0+\frac1{\sqrt{6}}\eta & K^0 \\    
~~K^- & \bar{K}^0 & -\frac2{\sqrt{6}}\eta
\end{array}             
\right) \, ,
\end{eqnarray}
and $m$ is the mass matrix expressed as
\begin{eqnarray}
m=\left(                
\begin{array}{ccc}                       
M_\pi^2 & 0  &0  \\    
~0 & M_\pi^2 &0 \\    
~0 & 0 & 2M_K^2-M_\pi^2
\end{array}             
\right)\, .
\end{eqnarray}
In our extrapolation, the next higher order Lagrangian will provide the leading tree level contribution.
The one-loop Feynman diagram for the magnetic polarizability of the charged pion is shown in Fig.~\ref{fig:Nfig1}.

\begin{figure}[t]
\begin{center}
\includegraphics[scale=0.85]{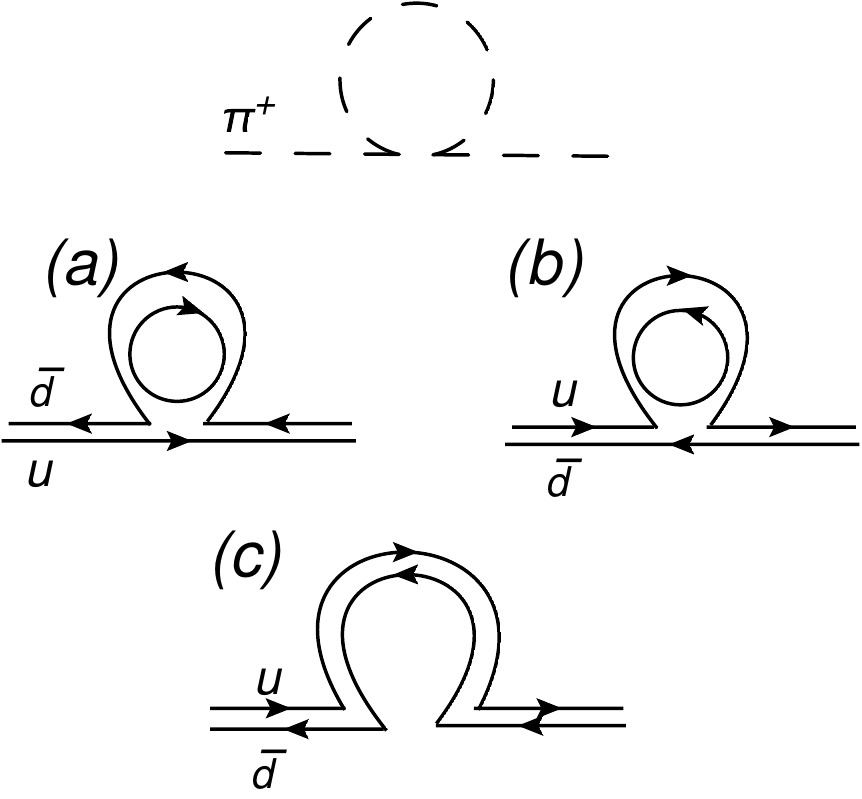}
\caption{Quark-flow diagrams for the $\pi^+\pi^+\pi^+\pi^+$ channel. The couplings for
  Figs.~2(a) and 2(b) can be obtained individually by replacing the light sea-quark-loop flavor with
  a strange quark flavor and using SU(3)-flavor symmetry \cite{Leinweber:2002qb}.}.
\label{fig:Nfig2}
\end{center}
\end{figure}

For the charged pion, the amplitude of Fig.~\ref{fig:Nfig1} is written as
\begin{widetext}
\begin{eqnarray}\label{eq:tot}
T=\frac{-ie^2}{3f_\pi^2}\int\,\frac{d^4k}{(2\pi)}\frac{[(k-q_1)(k-q_2)-3P(k-q_1)-3P(k-q_2)-M_\pi^2](2k-q_2)_\nu(2k-q_1)_\mu}{(k^2-M_\pi^2)((k-q_2)^2-M_\pi^2)((k-q_1)^2-M_\pi^2)}\epsilon^\mu(q_1)\epsilon^{\nu*}(q_2)+C.S.
\end{eqnarray}
\end{widetext}
where $C.S.$ denotes crossing symmetry terms where the photons labeled $q_{1,\mu}$ and $q_{2,\nu}$
in Fig.~\ref{fig:Nfig1} couple with the opposite time ordering. This amplitude has no contribution to the magnetic
polarizability of the charged pion. 

The lattice results in \cite{Bignell:2020dze} are simulated in the electroquenched approximation.
In this case, virtual sea-quark loops in the QCD vacuum do not interact with the background
magnetic field.  To understand the impact of this approximation, we draw on partially quenched
chiral perturbation theory to separate the contributions of sea-quark loops and understand their
role in composing the properties of QCD.  

Hu {\it et al.} \cite{Hu:2007ts} have performed a comprehensive calculation of the pion
polarizability at one loop using the graded-symmetry formalism of partial quenching for isolating
sea-quark-loop contributions \cite{Bernard:1992mk}.  Here we complement this approach with a simple
diagrammatic approach \cite{Leinweber:2002qb} where the sea-quark-loop contributions are isolated
by changing the sea-quark flavor to a flavor that does not appear in the hadron under
consideration, in this case a strange quark.  Drawing on established SU(3) flavor relations the
loop contribution is readily obtained.

The quark-flow diagrams for the $\pi^+\pi^+\pi^+\pi^+$ channel are plotted in Fig.~\ref{fig:Nfig2}.
The coefficients of the four-meson vertex for Figs.~\ref{fig:Nfig2}(a) and (b) alone can be obtained by
replacing the sea quark with a strange quark \cite{Leinweber:2002qb} as described above.
Therefore, the amplitude can be obtained by calculating a $K$-meson loop dressing with the
$K$-meson mass set equal to the pion mass.  The amplitude of Fig.~\ref{fig:Nfig2}(a) is written as
\begin{widetext}
\begin{eqnarray}\label{eq:ds1}
T^{(a)}=\frac{-ie^2}{6f_\pi^2}\int\,\frac{d^4k}{(2\pi)}\frac{[(k-q_1)(k-q_2)-3P(k-q_1)-3P(k-q_2)-M_\pi^2](2k-q_2)_\nu(2k-q_1)_\mu}{(k^2-M_\pi^2)((k-q_2)^2-M_\pi^2)((k-q_1)^2-M_\pi^2)}\epsilon^\mu(q_1)\epsilon^{\nu*}(q_2)+C.S. \, .
\end{eqnarray}
\end{widetext}
Except for the leading factor, Eq.~(\ref{eq:ds1}) is the same as Eq.~(\ref{eq:tot}) and once again
this amplitude does not contribute to the magnetic polarizability of the charged pion.  Since the
expression of Fig.~\ref{fig:Nfig2}(b) has the same structure, its contribution also vanishes.
Since the sum of the contributions of Figs.~\ref{fig:Nfig2}(a) and (b) matches the total amplitude
of Fig.~\ref{fig:Nfig1}, the $4\pi$ vertex in Fig.~\ref{fig:Nfig2}(c) also vanishes. These results
are consistent with the conclusions from graded symmetry \cite{Bernard:1992mk} and the
comprehensive analysis of Hu {\it et al.} \cite{Hu:2007ts}.

Thus, the one-loop diagrams of the leading order Lagrangian ${\mathcal L}_2$ make no contributions
to $\beta_{\pi^{\pm}}$ in either full QCD or electroquenched QCD. Loop contributions commence at
the two-loop level. This is in contrast with the $\pi^0$ case, where quark-annihilation
contractions of quark-field operators within the neutral-pion interpolating fields create
source-sink-disconnected contributions to $\beta_{\pi^0}$.  Similar contractions do not exist in
the charged pion interpolators and thus this one-loop amplitude does not appear for the
charged-pion magnetic polarizability.

The lowest order tree-level contribution starts from $\mathcal{L}_4$ and can be written as
\begin{equation}\label{eq:leading}
\beta_0=-\frac{\alpha}{16f_\pi^2M_\pi\pi^2}\frac{\bar{l}_\Delta}{3} \, ,
\end{equation}
where $\alpha=1/137$ is the fine structure constant, $\bar{l}_\Delta = 3.0\pm0.3$ is the
renormalized constant taken from Refs.~\cite{Burgi:1996qi,Gasser:2006qa}.  This is the leading
contribution to the magnetic polarizability of the charged pion, as it is proportional to
$1/M_\pi$. At the physical point, this contribution is $-2.98\,(30)\times 10^{-4}$ fm$^3$.
 
Note, for the $\pi^0$ case, the leading tree-level contribution is of order $M_\pi$. There,
sigma-meson exchange, contributing at order $1/M_\pi$, was included to describe the pion mass
dependence of the lattice results~\cite{He:2020ysm}. Here however, it is not necessary to introduce
sigma exchange as the $1/M_\pi$ contribution is effectively included in the $\bar{l}_\Delta$ term.

\section{Pad\'e Approximant}

The tree level contribution from $\mathcal{L}_6$ is $\mathcal{O}(M_\pi)$. Thus the tree-level
contribution up to $M_\pi^3$ can be written
\begin{equation}
\beta_{\rm tree}^{\pi^{\pm}} = \beta_0 + b_1 M_\pi + b_3 M_\pi^3=\beta_0(1 + c_1 M_\pi^2 + c_3
M_\pi^4) \, .
\end{equation} 
To create a function able to interpolate between the light quark-mass regime where $\chi$PT is
robust to the larger quark-mass regime where the lattice-QCD results are smooth and slowly varying,
we consider the following Pad\'e approximant
\begin{equation}\label{eq:ex0}
\beta^{\pi^\pm}_{L}=\beta_0\frac{1+c_1 M_\pi^2 + c_3 M_\pi^4}{1+c_4 M_\pi^4} \, ,
\end{equation}
where $c_1 = \frac{3a_1^r}{32f_\pi^2\pi^2\bar{l}_\Delta}$ and $a_1^r = -3.2$ is the renormalized low-energy constant as determined in
Ref.~\cite{Gasser:2006qa}. We note that the result for $\beta^{\pi^\pm}_L$ is insensitive to the value taken for $a_1^r$ as 
the fit parameters $c_3$ and $c_4$ compensate for changes in $a_1^r$ 
as the fit function is constrained by the lattice QCD results.

We note that the effectiveness of a Pad\'e in such circumstances was illustrated by a study of the exactly soluble Euler-Heisenberg effective action, where an approximant was shown to yield a surprisingly accurate fit to the exact result provided the logarithmic behavior at small mass and the inverse power behavior at large mass was correctly incorporated~\cite{Dunne:2001ip}. The parameters $c_3$ and $c_4$ are fit parameters constrained by fitting
the lattice QCD results.  We find
\begin{equation}\label{eq:para}
c_3=-0.59\pm 0.16~{\rm fm}^4, \quad\mbox{and}\quad c_4=1.50\pm0.15~{\rm fm}^4.
\end{equation}

\begin{figure}[t]
\begin{center}
\includegraphics[scale=0.85]{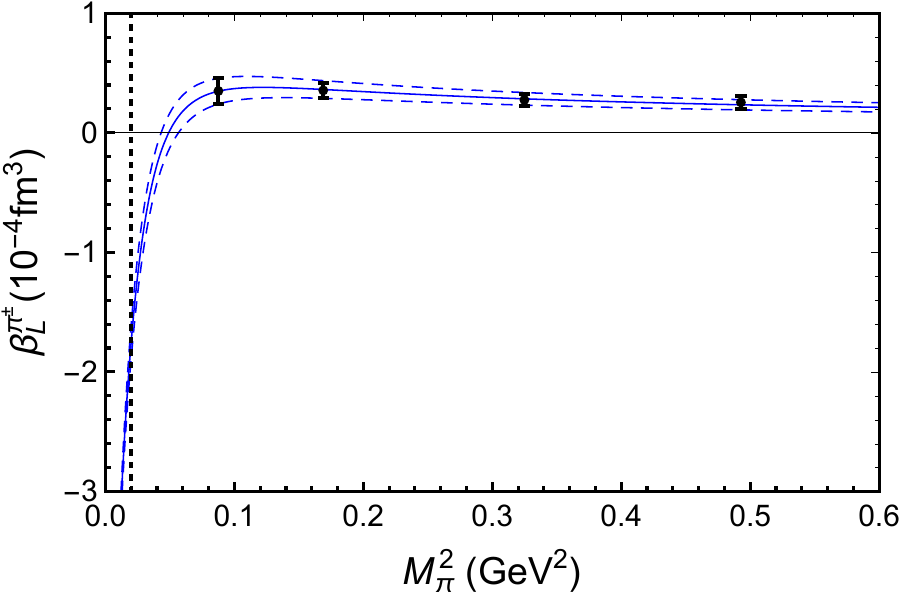}
\caption{Pion mass dependence of the magnetic polarizability of the charged pion.  The Pad\'e
  approximant of Eq.~(\ref{eq:ex0}) (solid curve) is constrained by the CSSM lattice results of
  Ref.~\cite{Bignell:2020dze} (black bullets) as described in the text.  The leading and
  next-to-leading contributions are constrained by $\chi$PT \cite{Burgi:1996qi,Gasser:2006qa}.
  Dashed curves represent the $1 \sigma$ error bounds associated with the statistical uncertainties
  of the lattice results.  The vertical dotted line indicates the physical pion-mass point.}
\label{fig:cp0}
\end{center}
\end{figure}

The pion mass dependence of the lattice QCD results for the charged-pion magnetic polarizability is
illustrated in Fig.~\ref{fig:cp0}.  The fit of $\beta_L^{\pi^\pm}$ to these lattice results is
illustrated as a solid curve.  One observes the lattice results can be described very well by the
Pad\'e approximant.  The Pad\'e approximant incorporates a heavy quark behavior consistent with
the lattice QCD observations, allowing the fit function to become flat at larger pion masses, 
which is of course not possible with a polynomial expansion.  At
small pion masses, $\beta_L^{\pi^\pm}$ is dominated by the results of $\chi$PT, decreasing quickly
with decreasing $M_\pi$ due to the leading order term proportional to $1/M_\pi$.  As a result, the
positive lattice results at large $M_\pi$ change to negative values at small $M_\pi$.  At the
physical pion mass, $\beta_{\pi^\pm}=-1.80(14)\times 10^{-4}$ fm$^3$.

This extrapolation is possible because there is knowledge of $\bar{l}_\Delta$ from chiral
perturbation theory. The merit of the extrapolation lies not only in providing a prediction to
confront experiment, but also in the guidance it provides for next generation lattice QCD
simulations to both directly observe the predicted sign change in the magnetic polarizability, and
determine the value of $\bar{l}_\Delta$ from the first principles of QCD.  To constrain
$\bar{l}_\Delta$, one needs precise lattice QCD results at small pion masses.  However, lattice
calculations have yet to resolve a signal there.  If chiral fermion actions are required to
circumvent difficulties associated with additive mass renormalization issues with Wilson-clover
fermions, a two-order of magnitude increase in computational resources will be required.

To investigate the importance of higher-order terms in the chiral expansion, we proceed to include
additional two-loop contributions from $\chi$PT \cite{Gasser:2006qa}.  In presenting these
contributions, we begin by simply adding the two-loop contributions to the existing fit illustrated
in Fig.~\ref{fig:cp0}.  The modified Pad\'e approximant incorporating the two-loop contributions
can be expressed as
\begin{equation}\label{eq:ex1}
\beta^{\pi^\pm}_F=\beta_0\frac{1 -\frac{3(d_{1+}-d_{1-})}{32\pi^2f^2_\pi \bar{l}_\Delta}M_\pi^2 + c_3 M_\pi^4}{1+c_4 M_\pi^4},
\end{equation}
where $d_{1+}$ and $d_{1-}$ are coefficients for the two-loop contributions \cite{Gasser:2006qa}. They are defined as
\begin{eqnarray}\label{eq:non-ana}
d_{1+}&=&8b^r-\frac49\Big\{l(l+\frac12\bar{l}_l+\frac32\bar{l}_2)-\frac{53}{24}l+\frac12\bar{l}_1+\frac32\bar{l}_2            \nonumber\\
&+&\frac{91}{72}+\Delta_+\Big\},  \nonumber\\
d_{1-}&=&a_1^r+8b^r-\frac43\Big\{l(\bar{l}_1-\bar{l}_2+\bar{l}_\Delta-\frac{65}{12})-\frac13\bar{l}_1-\frac13\bar{l}_2        \nonumber\\
&+&\frac14\bar{l}_3-\bar{l}_\Delta\bar{l}_4+\frac{187}{104}+\Delta_-\Big\},
\end{eqnarray} 
with 
\begin{eqnarray}
\Delta_+=\frac{8105}{576}-\frac{135}{64}\pi^2\, , \quad\mbox{and}\quad \Delta_-=\frac{41}{432}-\frac{53}{64}\pi^2\,,
\end{eqnarray}
where $l \equiv \ln \left ( {\mpi^2}/{\mu^2} \right ),$ and $\bar{l}_i$ are scale-independent low-energy
constants (LECs) defined in Eqs. (3.8), (3.9) and (3.10) of Ref.~\cite{Gasser:2006qa}. 
\begin{eqnarray}
\bar{l}_1&=&-0.4\pm0.6,~~~\bar{l}_2=4.3\pm0.1,   \nonumber\\
\bar{l}_3&=&2.9\pm2.4,~~~~~\bar{l}_4=4.4\pm0.2,    \nonumber\\
\bar{l}_\Delta&=&3.0\pm0.3.
\end{eqnarray}
The scale $\mu =
0.770$ GeV is set to the $\rho$-meson mass. 
The uncertainty in each of these LECs contributes to a systematic uncertainty in 
the magnetic polarizability $\beta_{\pi^\pm}$. We consider the uncertainty associated 
with each LEC and combine their contributions in quadrature.

\begin{figure}[t]
\begin{center}
\includegraphics[scale=0.85]{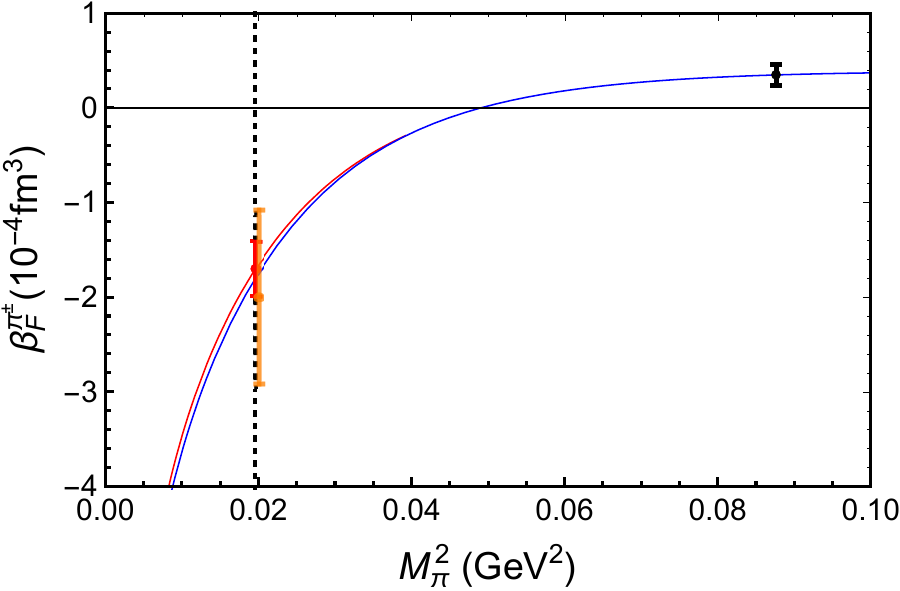}
\caption{Chiral extrapolation of the charged-pion magnetic polarizability.  The red and blue lines
  illustrate the Pad\'e approximates of Eqs.~(\ref{eq:ex1}) and (\ref{eq:ex0}) with and without the
  two-loop contributions respectively.  The full QCD prediction for the magnetic polarizability of
  the charged pion $\beta_{\pi^\pm}$ is illustrated at the physical pion mass by the red point
where the error bar incorporates both statistical and systematic uncertainties as
    described in the text.  The experimental measurement by the COMPASS collaboration
  \cite{Adolph:2014kgj} is illustrated by the orange point at the physical pion mass.}
\label{fig:cpe5}
\end{center}
\end{figure}

Our final results for the chiral extrapolation of the magnetic polarizability of the charged pion
is illustrated in Fig.~\ref{fig:cpe5}.  The blue line is our previous fit of Eq.~(\ref{eq:ex0}) to
the lattice results. The red line for $0 \le \mpi^2 \le 2\,\mpi^{2\,\rm Phys}$ represents the
chiral extrapolation with the two-loop contributions added to our previous fit as described by
Eq.~(\ref{eq:ex1}).  The addition of the two-loop contributions makes only a small adjustment to
the chiral extrapolation. At the physical point, the two-loop correction is
 $0.10 \times 10^{-4}$ fm$^3$, a 6\% correction.  However the correction decreases as
one moves to heavier pion masses.  Given that the corrections are very small and it is not clear to
what extent electroquenched simulations incorporate these effects, we propose the addition of
two-loop effects to provide the better estimate of the observable, and adopt the difference between
the red and blue curves as contributing to the systematic error, added in quadrature.

\begin{table*}[t] 
\caption{Comparison of the contributions to the charged
  pion magnetic polarizability in the standard units of $\times 10^{-4}$ fm$^3$.}
\label{tab:cont}
\begin{ruledtabular}
\begin{tabular}{lcc}
  Description  &Term  &Value ($\times 10^{-4}$ fm$^3$) \\
  \noalign{\smallskip}
  \hline
  \noalign{\smallskip}
  Total contribution without the two-loop correction   &Eq.~(\ref{eq:ex0})      & -1.80\\
    \noalign{\smallskip}
  \hline
   \noalign{\smallskip}
  Tree level contribution at leading order &  $\beta_0$ & -2.98\\
  \noalign{\smallskip}
  Correction $\propto M_\pi^2$ &$\displaystyle \displaystyle \beta_0c_1M_\pi^2$ & 0.07\\
   \noalign{\smallskip}
  Correction $\propto M_\pi^4$ &$\displaystyle \beta_0(c_3-c_4)M_\pi^4$ & 1.60\\
   \noalign{\smallskip}   
Sum of remaining contributions &$\beta_L^{\pi^{\pm}}-\beta_0(1+c_1M_\pi^2+(c_3-c_4)M_\pi^4)$ & -0.49\\
   \noalign{\smallskip}     
   \hline
   \noalign{\smallskip}
   Two loop correction of Eq.~(\ref{eq:ex1}) &$\displaystyle -\frac{3(d_{1+}-d_{1-}+a_1^r)}{32\pi^2f^2_\pi \bar{l}_\Delta(1+c_4 M_\pi^4)}M_\pi^2\beta_0 $     &0.10 \\
   \noalign{\smallskip}
   \hline
   \noalign{\smallskip}
   Full QCD prediction &Eq.~(\ref{eq:ex1}) & -1.70 \\ 
\end{tabular}
\end{ruledtabular}
\vspace{-12pt}
\end{table*}

The full QCD prediction for the magnetic polarizability of the charged pion $\beta_{\pi^\pm}$ is
illustrated at the physical pion mass in Fig.~\ref{fig:cpe5} by the red point where
the error bar incorporates both statistical and systematic uncertainties combined in quadrature.
Our final estimate is $\beta_{\pi^\pm}=-1.70\,(14)_{\rm stat}(25)_{\rm syst}\times 10^{-4}$ fm$^3$.

The experimental measurement obtained by the COMPASS collaboration \cite{Adolph:2014kgj} is
$\beta_{\pi^\pm}=(-2.0 \pm 0.6 \pm 0.7)\times 10^{-4}$ fm$^3$ under the assumption
$(\beta+\alpha)^\pi=0$.  Here the uncertainties are statistical and systematic, 
respectively.  This measurement is illustrated by the orange point in Fig.~\ref{fig:cpe5} where the
statistical and systematic uncertainties have been added in quadrature.  Our result is in good
agreement with the experimental measurement.  

It is interesting to examine how a power-series expansion of the Pad\'e generates corrections to
the leading contribution of $\beta_0$.
We commence with a Taylor expansion of the total contribution without the two-loop correction, in
Eq.~(\ref{eq:ex0})
\begin{equation}
\beta^{\pi^\pm}_L = \beta_0\left(1+c_1M_\pi^2 + (c_3-c_4) M_\pi^4\right)+\cdots \, .
\end{equation}
Table~\ref{tab:cont}, lists the contributions at different orders of the expansion evaluated at the
physical pion mass.  Recall that the coefficient $c_1$ is related to $a_1^r$, known from chiral
perturbation theory, whereas $c_3$ and $c_4$ are constrained by the current lattice QCD results.

We note that the large contribution at $\propto M_\pi^4$ may be a reflection of the Pad\'e
considered where a ratio of $c_3/c_4$ is encountered at large pion mass. For example, a Pad\'e
involving a ratio of $M_\pi^6$ terms may shift the strength observed at $M_\pi^4$ to neighboring
terms. On the other hand, these coefficients are required to describe the lattice QCD results.
Therefore, the large contribution at $M_\pi^4$ cautions against the naive application of the
power-series expansion, even at the physical pion mass.  

Considering the corrections $\propto M_\pi^4$ and higher order contributions together, we observe
an $\mathcal{O}(M_\pi^4)$ correction of $1.11 \times 10^{-4}$ fm$^3$ relative to a leading order
term of magnitude $2.98 \times 10^{-4}$ fm$^3$, a 37\% correction at the physical pion mass.

\vspace{-12pt}
\section{Summary}

In this paper, we have investigated the magnetic polarizability of the charged pion based on an
analysis of recent lattice QCD simulations at a range of quark masses.  We considered
partially-quenched chiral perturbation theory to understand the role of sea-quark-loops in the magnetic
polarizability of the charged pion at one-loop level, vital to understanding the impact of
electroquenching in the lattice QCD simulations.  In this case, electroquenched and full QCD
agree, with neither theory making contributions to the charged-pion magnetic polarizability at
one-loop level. Thus the fact that the lattice simulations are electroquenched has no impact on
the leading loop contributions to the magnetic polarizability.

To interpolate between the light quark-mass regime where $\chi$PT is robust to the larger
quark-mass regime where the lattice-QCD results are smooth and slowly varying, a Pad\'e
approximant was constructed.  The lattice results are described well by the Pad\'e approximant.
To evaluate the impact of higher-order contributions in the chiral extrapolation, two-loop
contributions as determined by Ref.~\cite{Gasser:2006qa} were investigated.  The contribution at
the physical pion mass is small and decreases as one moves to larger pion masses.

Although the lattice results at larger pion masses are positive, the final result at the physical
pion mass is negative at $\beta_{\pi^\pm}=-1.70\,(14)_{\rm stat}(25)_{\rm syst}\times 10^{-4}$
fm$^3$, in very good agreement with the experimental measurement by the COMPASS collaboration
\cite{Adolph:2014kgj}.

Future research will focus on studying the pion magnetic polarizability at smaller pion masses, to
directly observe the sign change of $\beta_{\pi^\pm}$ predicted in this analysis.  Such simulations
are very demanding, particularly if the  large statistical fluctuations observed at light quark
masses are associated with additive-mass renormalization issues in nonchiral fermion actions.
If one needs a chiral fermion action to circumvent this problem, a two-order of magnitude 
in computational resources will be required. 

Understanding the finite volume corrections to the charged pion magnetic polarizability remains of
interest, and can be quantified through simulations on larger volumes.

An alternative approach to the background field formalism is to access the polarizability via
perturbative electromagnetic-current insertions in four-point correlation function calculations
\cite{Wilcox:1996vx,Wilcox:2021rtt}.  Here the QCD basis states are not mixed by the
electromagnetic interactions and may be of advantage in understanding $\Sigma^0$ and $\Lambda$
polarizabilities for example, as these states mix in the background-field formalism.

Finally, while the two-loop contributions are remarkably small, in principle it would be
interesting to bring the graded-symmetry approach of partially-quenched chiral perturbation theory
to the two-loop $\chi$PT calculations to learn the details of electroquenching in the two-loop
sector.

\vspace{-18pt}
\section*{Acknowledgement}
\vspace{-6pt}
This research was supported with supercomputing resources provided by the Phoenix HPC service at
the University of Adelaide. This research was undertaken with the assistance of resources from the
National Computational Infrastructure (NCI), provided through the National Computational Merit
Allocation Scheme, and supported by the Australian Government through Grants No.~LE190100021,
LE160100051 and the University of Adelaide Partner Share.  This research was supported by the
Australian Research Council through ARC Discovery Project Grants No. DP180100497
(A.W.T) and DP150103164, DP190102215 and DP210103706 (D.B.L), and by the National Natural Sciences
Foundations of China under the grant No. 11975241.

\bibliographystyle{utphys2}
\bibliography{ref}

\providecommand{\href}[2]{#2}\begingroup\raggedright\begin{thebibliography}{10}

\bibitem{Ahrens:2004mg}
J.~Ahrens {\em et~al.} \href{http://dx.doi.org/10.1140/epja/i2004-10056-2}{{\em
  Eur. Phys. J.} {\bfseries A23} (2005) 113--127},
\href{http://arxiv.org/abs/nucl-ex/0407011}{{\ttfamily arXiv:nucl-ex/0407011
  [nucl-ex]}}.

\bibitem{Adolph:2014kgj}
{\bfseries COMPASS} Collaboration, C.~Adolph {\em et~al.}
  \href{http://dx.doi.org/10.1103/PhysRevLett.114.062002}{{\em Phys. Rev.
  Lett.} {\bfseries 114} (2015) 062002},
\href{http://arxiv.org/abs/1405.6377}{{\ttfamily arXiv:1405.6377 [hep-ex]}}.

\bibitem{Ivanov:1991kw}
M.~A. Ivanov and T.~Mizutani
\href{http://dx.doi.org/10.1103/PhysRevD.45.1580}{{\em Phys. Rev.} {\bfseries
  D45} (1992) 1580--1601}.

\bibitem{Bernard:1988wi}
V.~Bernard and D.~Vautherin
\href{http://dx.doi.org/10.1103/PhysRevD.40.1615}{{\em Phys. Rev.} {\bfseries
  D40} (1989) 1615}.

\bibitem{Dorokhov:1997rv}
A.~E. Dorokhov, M.~K. Volkov, J.~Hufner, S.~P. Klevansky, and P.~Rehberg
\href{http://dx.doi.org/10.1007/s002880050454}{{\em Z. Phys.} {\bfseries C75}
  (1997) 127--135}.

\bibitem{Bernard:1988gp}
V.~Bernard, B.~Hiller, and W.~Weise
\href{http://dx.doi.org/10.1016/0370-2693(88)90391-7}{{\em Phys. Lett.}
  {\bfseries B205} (1988) 16--21}.

\bibitem{Filkov:1982cx}
L.~V. Filkov, I.~Guiasu, and E.~E. Radescu
\href{http://dx.doi.org/10.1103/PhysRevD.26.3146}{{\em Phys. Rev.} {\bfseries
  D26} (1982) 3146}.

\bibitem{Donoghue:1993kw}
J.~F. Donoghue and B.~R. Holstein
  \href{http://dx.doi.org/10.1103/PhysRevD.48.137}{{\em Phys. Rev.} {\bfseries
  D48} (1993) 137--146},
\href{http://arxiv.org/abs/hep-ph/9302203}{{\ttfamily arXiv:hep-ph/9302203
  [hep-ph]}}.

\bibitem{Burgi:1996qi}
U.~Burgi \href{http://dx.doi.org/10.1016/0550-3213(96)00454-3}{{\em Nucl.
  Phys.} {\bfseries B479} (1996) 392--426},
\href{http://arxiv.org/abs/hep-ph/9602429}{{\ttfamily arXiv:hep-ph/9602429
  [hep-ph]}}.

\bibitem{Gasser:2006qa}
J.~Gasser, M.~A. Ivanov, and M.~E. Sainio
  \href{http://dx.doi.org/10.1016/j.nuclphysb.2006.03.022}{{\em Nucl. Phys.}
  {\bfseries B745} (2006) 84--108},
\href{http://arxiv.org/abs/hep-ph/0602234}{{\ttfamily arXiv:hep-ph/0602234
  [hep-ph]}}.

\bibitem{Martinelli:1982cb}
G.~Martinelli, G.~Parisi, R.~Petronzio, and F.~Rapuano
  \href{http://dx.doi.org/10.1016/0370-2693(82)90162-9}{{\em Phys. Lett. B}
  {\bfseries 116} (1982) 434--436}.

\bibitem{Bernard:1982yu}
C.~W. Bernard, T.~Draper, K.~Olynyk, and M.~Rushton
  \href{http://dx.doi.org/10.1103/PhysRevLett.49.1076}{{\em Phys. Rev. Lett.}
  {\bfseries 49} (1982) 1076}.

\bibitem{Fiebig:1988en}
H.~R. Fiebig, W.~Wilcox, and R.~M. Woloshyn
  \href{http://dx.doi.org/10.1016/0550-3213(89)90180-6}{{\em Nucl. Phys. B}
  {\bfseries 324} (1989) 47--66}.

\bibitem{Burkardt:1996vb}
M.~Burkardt, D.~B. Leinweber, and X.-m. Jin
  \href{http://dx.doi.org/10.1016/0370-2693(96)00881-7}{{\em Phys. Lett.}
  {\bfseries B385} (1996) 52--56},
\href{http://arxiv.org/abs/hep-ph/9604450}{{\ttfamily arXiv:hep-ph/9604450
  [hep-ph]}}.

\bibitem{Lee:2005dq}
F.~X. Lee, L.~Zhou, W.~Wilcox, and J.~C. Christensen
  \href{http://dx.doi.org/10.1103/PhysRevD.73.034503}{{\em Phys. Rev. D}
  {\bfseries 73} (2006) 034503},
  \href{http://arxiv.org/abs/hep-lat/0509065}{{\ttfamily
  arXiv:hep-lat/0509065}}.

\bibitem{Primer:2013pva}
T.~Primer, W.~Kamleh, D.~Leinweber, and M.~Burkardt
  \href{http://dx.doi.org/10.1103/PhysRevD.89.034508}{{\em Phys. Rev. D}
  {\bfseries 89} no.~3, (2014) 034508},
  \href{http://arxiv.org/abs/1307.1509}{{\ttfamily arXiv:1307.1509 [hep-lat]}}.

\bibitem{Luschevskaya:2014lga}
E.~Luschevskaya, O.~Solovjeva, O.~Kochetkov, and O.~Teryaev
  \href{http://dx.doi.org/10.1016/j.nuclphysb.2015.07.023}{{\em Nucl. Phys. B}
  {\bfseries 898} (2015) 627--643},
  \href{http://arxiv.org/abs/1411.4284}{{\ttfamily arXiv:1411.4284 [hep-lat]}}.

\bibitem{Luschevskaya:2015cko}
E.~Luschevskaya, O.~Solovjeva, and O.~Teryaev
  \href{http://dx.doi.org/10.1016/j.physletb.2016.08.054}{{\em Phys. Lett. B}
  {\bfseries 761} (2016) 393--398},
  \href{http://arxiv.org/abs/1511.09316}{{\ttfamily arXiv:1511.09316
  [hep-lat]}}.

\bibitem{Bali:2017ian}
G.~S. Bali, B.~B. Brandt, G.~Endrődi, and B.~Gläßle
  \href{http://dx.doi.org/10.1103/PhysRevD.97.034505}{{\em Phys. Rev.}
  {\bfseries D97} no.~3, (2018) 034505},
\href{http://arxiv.org/abs/1707.05600}{{\ttfamily arXiv:1707.05600 [hep-lat]}}.

\bibitem{Bignell:2019vpy}
R.~Bignell, W.~Kamleh, and D.~Leinweber
  \href{http://dx.doi.org/10.1103/PhysRevD.100.114518}{{\em Phys. Rev.}
  {\bfseries D100} no.~11, (2020) 114518},
  \href{http://arxiv.org/abs/1910.14244}{{\ttfamily arXiv:1910.14244
  [hep-lat]}}.
[Phys. Rev.D100,114518(2019)].

\bibitem{Ding:2020hxw}
H.-T. Ding, S.-T. Li, A.~Tomiya, X.-D. Wang, and Y.~Zhang
  \href{http://arxiv.org/abs/2008.00493}{{\ttfamily arXiv:2008.00493
  [hep-lat]}}.

\bibitem{Chang:2015qxa}
{\bfseries NPLQCD} Collaboration, E.~Chang, W.~Detmold, K.~Orginos, A.~Parreno,
  M.~J. Savage, B.~C. Tiburzi, and S.~R. Beane
  \href{http://dx.doi.org/10.1103/PhysRevD.92.114502}{{\em Phys. Rev. D}
  {\bfseries 92} no.~11, (2015) 114502},
  \href{http://arxiv.org/abs/1506.05518}{{\ttfamily arXiv:1506.05518
  [hep-lat]}}.

\bibitem{Bignell:2018acn}
R.~Bignell, J.~Hall, W.~Kamleh, D.~Leinweber, and M.~Burkardt
  \href{http://dx.doi.org/10.1103/PhysRevD.98.034504}{{\em Phys. Rev.}
  {\bfseries D98} no.~3, (2018) 034504},
\href{http://arxiv.org/abs/1804.06574}{{\ttfamily arXiv:1804.06574 [hep-lat]}}.

\bibitem{Bignell:2020xkf}
R.~Bignell, W.~Kamleh, and D.~Leinweber
  \href{http://dx.doi.org/10.1103/PhysRevD.101.094502}{{\em Phys. Rev.}
  {\bfseries D101} no.~9, (2020) 094502},
\href{http://arxiv.org/abs/2002.07915}{{\ttfamily arXiv:2002.07915 [hep-lat]}}.

\bibitem{Bignell:2020dze}
R.~Bignell, W.~Kamleh, and D.~Leinweber
  \href{http://dx.doi.org/10.1016/j.physletb.2020.135853}{{\em Phys. Lett.}
  {\bfseries B811} (2020) 135853},
\href{http://arxiv.org/abs/2005.10453}{{\ttfamily arXiv:2005.10453 [hep-lat]}}.

\bibitem{Hall:2013dva}
J.~M.~M. Hall, D.~B. Leinweber, and R.~D. Young
  \href{http://dx.doi.org/10.1103/PhysRevD.89.054511}{{\em Phys. Rev.}
  {\bfseries D89} no.~5, (2014) 054511},
\href{http://arxiv.org/abs/1312.5781}{{\ttfamily arXiv:1312.5781 [hep-lat]}}.

\bibitem{He:2020ysm}
F.~He, D.~B. Leinweber, A.~W. Thomas, and P.~Wang
  \href{http://dx.doi.org/10.1103/PhysRevD.102.114509}{{\em Phys. Rev.}
  {\bfseries D102} no.~11, (2020) 114509},
\href{http://arxiv.org/abs/2010.01580}{{\ttfamily arXiv:2010.01580 [nucl-th]}}.

\bibitem{Thomas:2002sj}
A.~W. Thomas \href{http://dx.doi.org/10.1016/S0920-5632(03)01492-0}{{\em Nucl.
  Phys. Proc. Suppl.} {\bfseries 119} (2003) 50--58},
\href{http://arxiv.org/abs/hep-lat/0208023}{{\ttfamily arXiv:hep-lat/0208023
  [hep-lat]}}.

\bibitem{Leinweber:2003dg}
D.~B. Leinweber, A.~W. Thomas, and R.~D. Young
  \href{http://dx.doi.org/10.1103/PhysRevLett.92.242002}{{\em Phys. Rev. Lett.}
  {\bfseries 92} (2004) 242002},
\href{http://arxiv.org/abs/hep-lat/0302020}{{\ttfamily arXiv:hep-lat/0302020
  [hep-lat]}}.

\bibitem{Leinweber:2005xz}
D.~B. Leinweber, A.~W. Thomas, and R.~D. Young
  \href{http://dx.doi.org/10.1016/j.nuclphysa.2005.03.024}{{\em Nucl. Phys.}
  {\bfseries A755} (2005) 59--70},
\href{http://arxiv.org/abs/hep-lat/0501028}{{\ttfamily arXiv:hep-lat/0501028
  [hep-lat]}}.

\bibitem{Allton:2005fb}
C.~Allton, W.~Armour, D.~B. Leinweber, A.~W. Thomas, and R.~D. Young
  \href{http://dx.doi.org/10.1016/j.physletb.2005.09.020}{{\em Phys. Lett. B}
  {\bfseries 628} (2005) 125--130},
  \href{http://arxiv.org/abs/hep-lat/0504022}{{\ttfamily
  arXiv:hep-lat/0504022}}.

\bibitem{Armour:2005mk}
W.~Armour, C.~Allton, D.~B. Leinweber, A.~W. Thomas, and R.~D. Young
  \href{http://dx.doi.org/10.1088/0954-3899/32/7/007}{{\em J. Phys. G}
  {\bfseries 32} (2006) 971--992},
  \href{http://arxiv.org/abs/hep-lat/0510078}{{\ttfamily
  arXiv:hep-lat/0510078}}.

\bibitem{Young:2004tb}
R.~D. Young, D.~B. Leinweber, and A.~W. Thomas
  \href{http://dx.doi.org/10.1103/PhysRevD.71.014001}{{\em Phys. Rev.}
  {\bfseries D71} (2005) 014001},
\href{http://arxiv.org/abs/hep-lat/0406001}{{\ttfamily arXiv:hep-lat/0406001
  [hep-lat]}}.

\bibitem{Leinweber:2004tc}
D.~B. Leinweber, S.~Boinepalli, I.~C. Cloet, A.~W. Thomas, A.~G. Williams,
  R.~D. Young, J.~M. Zanotti, and J.~B. Zhang
  \href{http://dx.doi.org/10.1103/PhysRevLett.94.212001}{{\em Phys. Rev. Lett.}
  {\bfseries 94} (2005) 212001},
\href{http://arxiv.org/abs/hep-lat/0406002}{{\ttfamily arXiv:hep-lat/0406002
  [hep-lat]}}.

\bibitem{Wang:1900ta}
P.~Wang, D.~Leinweber, A.~Thomas, and R.~Young
  \href{http://dx.doi.org/10.1103/PhysRevC.79.065202}{{\em Phys. Rev. C}
  {\bfseries 79} (2009) 065202},
  \href{http://arxiv.org/abs/0807.0944}{{\ttfamily arXiv:0807.0944 [hep-ph]}}.

\bibitem{Leinweber:2006ug}
D.~B. Leinweber, S.~Boinepalli, A.~W. Thomas, P.~Wang, A.~G. Williams, R.~D.
  Young, J.~M. Zanotti, and J.~B. Zhang
  \href{http://dx.doi.org/10.1103/PhysRevLett.97.022001}{{\em Phys. Rev. Lett.}
  {\bfseries 97} (2006) 022001},
\href{http://arxiv.org/abs/hep-lat/0601025}{{\ttfamily arXiv:hep-lat/0601025
  [hep-lat]}}.

\bibitem{Hall:2013oga}
J.~M.~M. Hall, D.~B. Leinweber, and R.~D. Young
  \href{http://dx.doi.org/10.1103/PhysRevD.88.014504}{{\em Phys. Rev.}
  {\bfseries D88} no.~1, (2013) 014504},
\href{http://arxiv.org/abs/1305.3984}{{\ttfamily arXiv:1305.3984 [hep-lat]}}.

\bibitem{HackettJones:2000qk}
E.~J. Hackett-Jones, D.~B. Leinweber, and A.~W. Thomas
  \href{http://dx.doi.org/10.1016/S0370-2693(00)00899-6}{{\em Phys. Lett.}
  {\bfseries B489} (2000) 143--147},
\href{http://arxiv.org/abs/hep-lat/0004006}{{\ttfamily arXiv:hep-lat/0004006
  [hep-lat]}}.

\bibitem{Bijnens:1987dc}
J.~Bijnens and F.~Cornet
\href{http://dx.doi.org/10.1016/0550-3213(88)90032-6}{{\em Nucl. Phys.}
  {\bfseries B296} (1988) 557--568}.

\bibitem{Groom:2000in}
{\bfseries Particle Data Group} Collaboration, D.~E. Groom {\em et~al.}
{\em Eur. Phys. J.} {\bfseries C15} (2000) 1--878.

\bibitem{Leinweber:2002qb}
D.~B. Leinweber \href{http://dx.doi.org/10.1103/PhysRevD.69.014005}{{\em Phys.
  Rev.} {\bfseries D69} (2004) 014005},
\href{http://arxiv.org/abs/hep-lat/0211017}{{\ttfamily arXiv:hep-lat/0211017
  [hep-lat]}}.

\bibitem{Hu:2007ts}
J.~Hu, F.-J. Jiang, and B.~C. Tiburzi
  \href{http://dx.doi.org/10.1103/PhysRevD.77.014502}{{\em Phys. Rev.}
  {\bfseries D77} (2008) 014502},
\href{http://arxiv.org/abs/0709.1955}{{\ttfamily arXiv:0709.1955 [hep-lat]}}.

\bibitem{Bernard:1992mk}
C.~W. Bernard and M.~F.~L. Golterman
  \href{http://dx.doi.org/10.1103/PhysRevD.46.853}{{\em Phys. Rev.} {\bfseries
  D46} (1992) 853--857},
\href{http://arxiv.org/abs/hep-lat/9204007}{{\ttfamily arXiv:hep-lat/9204007
  [hep-lat]}}.

\bibitem{Dunne:2001ip}
G.~V. Dunne, A.~W. Thomas, and S.~V. Wright
  \href{http://dx.doi.org/10.1016/S0370-2693(02)01363-1}{{\em Phys. Lett. B}
  {\bfseries 531} (2002) 77--82},
  \href{http://arxiv.org/abs/hep-th/0110155}{{\ttfamily arXiv:hep-th/0110155}}.

\bibitem{Wilcox:1996vx}
W.~Wilcox \href{http://dx.doi.org/10.1006/aphy.1996.5649}{{\em Annals Phys.}
  {\bfseries 255} (1997) 60--74},
  \href{http://arxiv.org/abs/hep-lat/9606019}{{\ttfamily
  arXiv:hep-lat/9606019}}.

\bibitem{Wilcox:2021rtt}
W.~Wilcox and F.~X. Lee \href{http://arxiv.org/abs/2106.02557}{{\ttfamily
  arXiv:2106.02557 [hep-lat]}}.

\end{thebibliography}\endgroup

\end{document}